\documentclass{aa}

\usepackage{amsmath,subcaption}
\usepackage[colorlinks=true,     linkcolor=blue, citecolor=blue, filecolor=blue, urlcolor=blue]{hyperref}

\begin{document}

\title{Apsidal Precession in Binary Asteroids}

\author{A. J. Meyer\inst{\ref{inst1}} \and D. J. Scheeres\inst{\ref{inst1}}}
\institute{Smead Department of Aerospace Engineering, University of Colorado Boulder, 3775 Discover Dr, Boulder, CO 80303, USA\label{inst1}}

\maketitle

\begin{abstract}

While the secondary in a binary asteroid plays an important role in the precession of the mutual orbit, this role has not been thoroughly studied. Given the complex spin-orbit coupled dynamics in binary asteroids, we use a numerical approach to study the relationship between the secondary's shape and spin and the apsidal precession rate of the orbit. Using this approach in conjunction with observations of Didymos, we find it is likely that Dimorphos was significantly reshaped as a result of the DART impact, with its new shape more elongated than the pre-impact shape. Finally, we show that non-principal axis rotation of the secondary can lead to a chaotic evolution of the longitude of the periapsis.

\end{abstract}

\section{Introduction}

The secular precession rate of an orbit is an important quantity driven by the physical characteristics of the system. It is well understood that the oblateness of the primary body plays a major role in the precession rate \citep{murray1999solar}. This fact has been used to estimate mass distributions within the body given an observed orbital precession rate. However, the typical assumption in these problems is that the secondary body is small and massless in comparison to the primary. This is not the case for binary asteroids, in which the secondary can be a significant fraction of the total mass in the system. Furthermore, the secondary is typically elongated, and its non-spherical shape can have an important influence on the system dynamics, especially in a perturbed system \citep{meyer2023perturbed}.

The topic of orbital precession in systems with irregular satellites has been previously discussed in the literature. \cite{borderies1990phobos} examine how the shape of Phobos influences its orbit and libration. However, this work does not account for free libration in the system, instead focusing only on the libration caused by the orbit's eccentricity (i.e., forced libration). Furthermore, Phobos's mass fraction is very small, being around 7 orders of magnitude smaller than Mars.

Specific to binary asteroids, \cite{fahnestock2008simulation} develop a relationship between secondary shape and orbit precession in their analysis of the dynamics of (66391) Moshup (known as 1999 KW4 at the time of their publication). However, several assumptions were made in the derivation of this expression that do not hold in general. \cite{meyer2023perturbed} showed that for librating, elongated secondaries the expression is inaccurate.

\cite{meyer2021libration} explored how the secondary shape can affect the observed orbit period in a binary asteroid, which experiences fluctuations due to the precession rate. This begins to connect the physical parameters of the secondary to the precession rate, but does not explicitly study this relationship. We take a more formal approach to understanding how the secondary parameters are manifested in observations of the precession rate.

Perhaps the best understanding of the secondary's effect on the precession rate is given by \cite{cuk2010orbital}. While this work explicitly relates the apsidal precession rate to the physical parameters of the secondary, it does not fully investigate how observations can constrain estimates of the secondary shape and dynamics. In the current work, we use the relationship defined in \cite{cuk2010orbital} to inform a numerical investigation on relating observations to physical and dynamical properties of the secondary. 

The primary motivation for this study is the recent DART impact. The DART impact into Dimorphos, the secondary in the Didymos binary asteroid system, perturbed the orbit resulting in an apsidal precession signal in observations. The estimated precession rate cannot be explained by the primary's oblateness alone, and after performing a formal fit on the Didymos data, \cite{Naidu2024} find a more elongated Dimorphos can explain the observed precession rate. In this work, we perform a deeper study of the role played by the secondary in the orbital precession rate within a binary asteroid. We then use this relationship to place constraints on the physical and dynamical properties of Dimorphos.

Additional analysis of the Didymos data suggests that Dimorphos is currently in a state of non-principal axis (NPA) rotation \citep{Pravec2024}. Due to the spin-orbit coupled nature of binary asteroid dynamics, this rotation state will have an effect on the orbit's precession rate. We will also investigate this relationship.

In the present study we first outline the dynamical model we will use in Section \ref{sec:model}, then summarize the analytic work done by \cite{cuk2010orbital} in Section \ref{sec:analytical}. We then present our general numerical study in Section \ref{sec:numerical}. Section \ref{sec:didymos} applies this approach to Didymos to place constraints on the shape and libration amplitude of Dimorphos. In Section \ref{sec:npa} we explore how the presence of NPA rotation changes the precession rate. Finally, we provide a discussion in Section \ref{sec:discussion} and a conclusion in Section \ref{sec:conclusion}.

\section{Dynamical Model}
\label{sec:model}
Since we are interested in the role played by the elongation and spin of the secondary in the precession rate of binary asteroids, we choose the simplest dynamical model that preserves those quantities. This is the sphere-ellipsoid model, where the primary is modeled as a sphere and the secondary as a triaxial ellipsoid. This will allow us to investigate the effects of only the secondary's shape and dynamics on the precession, eliminating the well-understood role played by the primary's oblateness.

\begin{figure}[ht!]
   \centering
   \includegraphics[width = 3in]{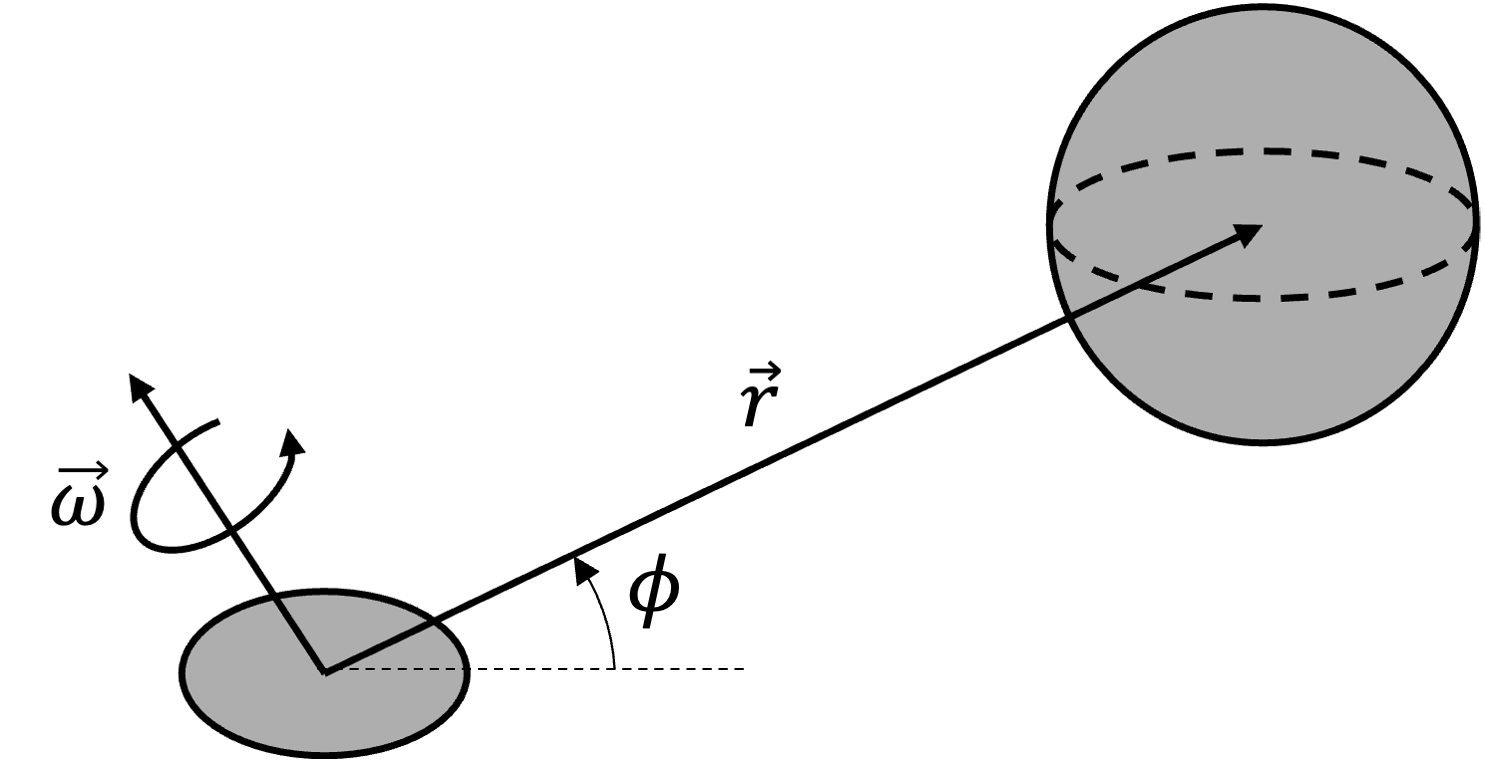} 
   \caption{A diagram of our dynamic model, the sphere-restricted full two-body problem. This shows the relative position and angular velocity vectors, centered in the secondary's body-fixed frame. The libration angle is also labeled.}
   \label{fig:dynamics}
\end{figure}

We show a diagram of our dynamical model in Fig. \ref{fig:dynamics}, showing the prolate secondary and spherical primary. This model permits full three-dimensional rotation and translation. The equations of motion for the sphere-restricted full two-body problem were derived in \cite{scheeres2006relative}, which take the secondary as the central body:

\begin{equation}
    \ddot{\vec{r}}+2\vec{\omega}\times\dot{\vec{r}}+\dot{\vec{\omega}}\times\vec{r}+\vec{\omega}\times(\vec{\omega}\times\vec{r})=\mathcal{G}(m_1+m_2)\frac{\partial U}{\partial \vec{r}}
    \label{eom1}
\end{equation}
\begin{equation}
    \mathbf{I}\cdot\dot{\vec{\omega}}+\vec{\omega}\times\mathbf{I}\cdot\vec{\omega}=-\mathcal{G}m_1m_2\vec{r}\times\frac{\partial U}{\partial \vec{r}}.
    \label{eom2}
\end{equation}
Here, $\vec{r}$ is the position of the primary relative to the secondary and $\vec{\omega}$ is the spin rate of the secondary. $\mathcal{G}$ is the universal gravitational constant, $m_1$ and $m_2$ are the mass of the primary and secondary, respectively. $U$ is the gravitational potential; with the secondary modeled as a triaxial ellipsoid we calculate the second-order potential using MacCullagh's formula \citep{murray1999solar}. The libration angle is an important quantity, which we define as the angle between the secondary's long axis and the relative position of the primary in the secondary's body-fixed frame. This is labeled as $\phi$ on Fig. \ref{fig:dynamics}.

We note the equations of motion, written in Eqs. \ref{eom1} and \ref{eom2}, are written in the body-fixed frame of the secondary. However, we are interested in the apsidal precession rate of the orbit, which is defined relative to an inertial frame. To account for this, we also integrate a simple set of quaternions in parallel with the equations of motion, allowing us to track the secondary's attitude relative to an inertial frame and perform a coordinate transformation between these two frames. The quaternions have the added benefit of having time derivatives which are a linear function of the spin vector \citep{schaub2003analytical}.

\section{Analytical Approach} \label{sec:analytical}

The groundwork for an analytic expression of the apsidal precession rate around an elongated body was done by \cite{burns1976elementary,burns1977erratum} through an application of the classical Gauss Planetary Equations. The radial and tangential accelerations due to the perturbation of the elongated body were derived in \cite{scheeres2009stability}. Using these, the time derivative of the longitude of periapsis, or equivalently the apsidal precession rate, as applied to binary asteroids was then finalized in \cite{cuk2010orbital}. The apsidal precession resulting from the body's oblateness is a classical result \citep{murray1999solar}, so here we focus only on the effect of the body's prolateness. For convenience we repeat the expression here, with corrections to minor errors in the original derivation. We also point out this equation is only defined for planar motion. Our numerical model allows for out-of-plane motion as well, so this expression should only be used if the dynamics are guaranteed to remain planar. The equation is non-dimensionalized, using the radius of the primary as the unit of length and the total system mass as the mass unit, leading to the unit of time equal to $1/2\pi$ of the orbit period at the equator of the primary \citep{cuk2010orbital}:

\begin{multline}
    \dot{\varpi} = \frac{\sqrt{a(1-e^2)}}{e}\frac{B-A}{r^4} \bigg[ \frac{9}{4}\cos 2\phi \cos f \\+ \frac{3}{2}\sin 2\phi \sin f \frac{2+e\cos f}{1+e\sin f} \bigg].
    \label{eq:cuk}
\end{multline}

Apart from the classical precession rate driven by oblateness, this expression shows the driving quantities in apsidal precession. Beyond the semimajor axis, $a$, eccentricity $e$, and separation $r$, we also see the precession rate depends on the $C_{22}$ gravity term (here equivalently written as $B-A$, where $A<B<C$ are the mass-normalized principal moments of inertia), the libration angle $\phi$, and the orbital position, measured by true anomaly $f$. Here we can directly observe increasing the secondary's $C_{22}$ increases the apsidal precession rate. The libration angle is the angle between the longest axis and the position vector, also written mathematically as:
\begin{equation}
    \phi=\gamma-(f+\varpi)
\end{equation}
where $\gamma$ is the elongated body's orientation in inertial space and $\varpi$ is the longitude of periapsis.

Interestingly, Eq. \ref{eq:cuk} includes the relative orientation of the prolate body. This orientation will depend on the spin state of the body, meaning perturbations to the body's spin will have an effect on the precession rate. This can be used to our advantage; if the precession rate of the orbit is measured, it can help constrain the spin rate and libration amplitude of the secondary in a binary asteroid. Here, we define our libration amplitude as the maximum of the libration angle.

Another point of interest is the role the true anomaly plays in the precession rate. \cite{scheeres2009stability} established that in the synchronous equilibrium of binary asteroids, the true anomaly is equal to zero and the secondary is trapped at periapsis, with the longitude of periapsis precessing at the orbit rate. This issue can be circumvented by using a set of orbit elements that account for the primary's oblateness, for example the geometric elements \citep{borderies1994test}. Within the Keplerian elements, this phenomenon is also reflected in Eq. \ref{eq:cuk}, where if the true anomaly and libration angle are equal to zero the apsidal precession rate is rapid. For small perturbations, the true anomaly and libration angle both oscillate about $0^\circ$. In this case, the precession rate is not constant, but is still rapid, since the longitude of periapsis is still circulating while the true anomaly oscillates. The critical point where the true anomaly switches from oscillation to circulation was derived in \cite{meyer2023perturbed}, who also briefly discuss the relationship between the true anomaly and apsidal precession rate.

\begin{figure*}
    \begin{subfigure}[t]{.5\textwidth}
    \centering
    \includegraphics[width=\linewidth]{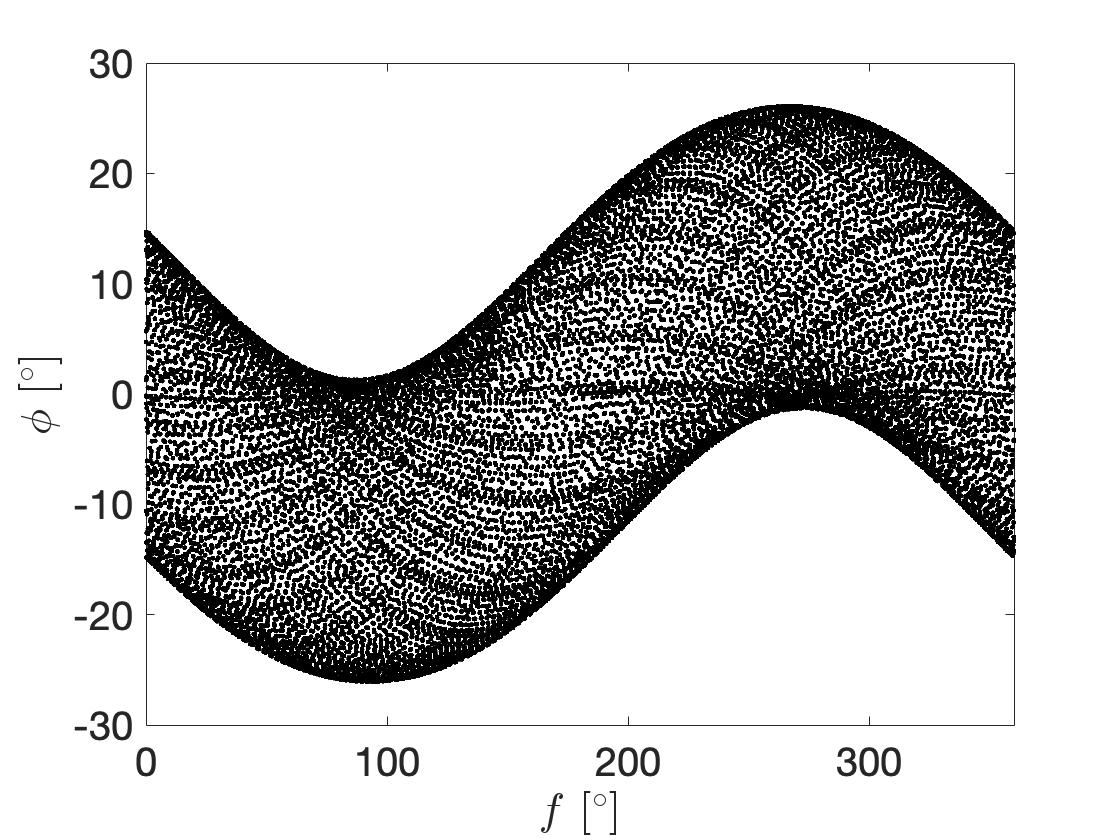}
    \caption{Libration as a function of true anomaly.}
    \end{subfigure}
    \hfill
    \begin{subfigure}[t]{.5\textwidth}
    \centering
    \includegraphics[width=\linewidth]{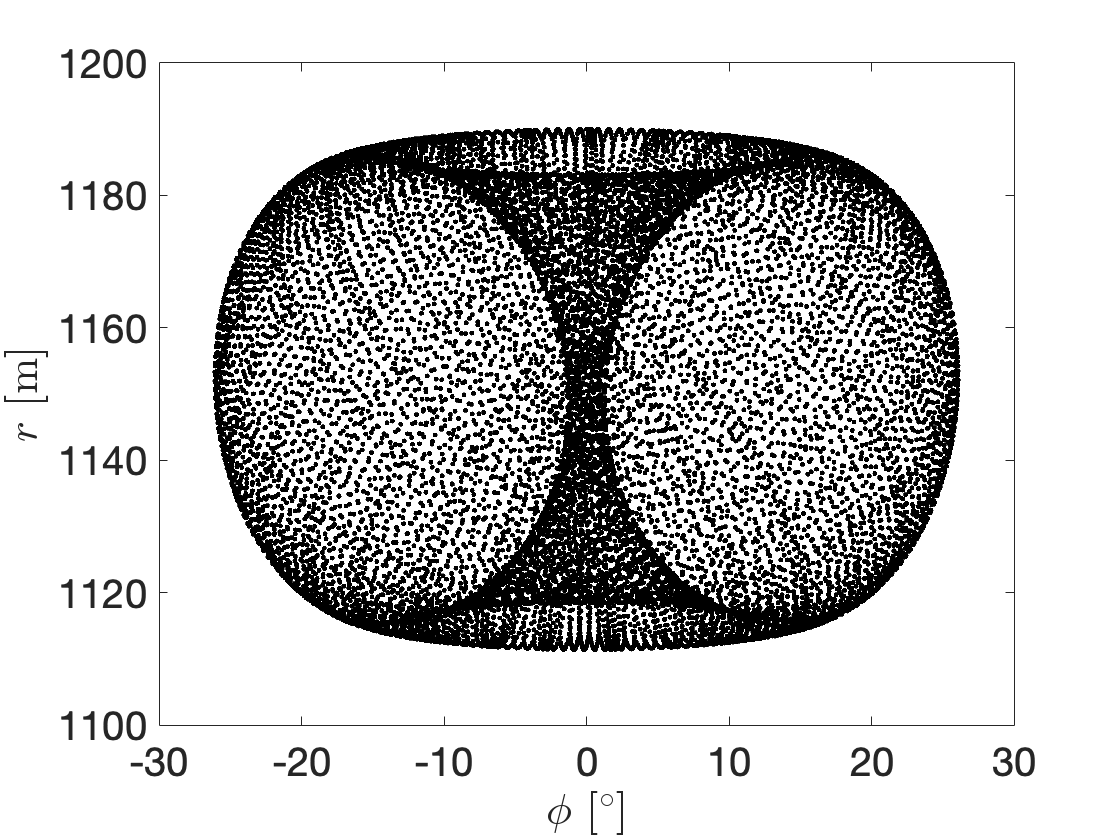}
    \caption{Separation as a function of libration.}
    \end{subfigure}
    
    \medskip
    
    \begin{subfigure}[t]{.5\textwidth}
    \centering
    \includegraphics[width=\linewidth]{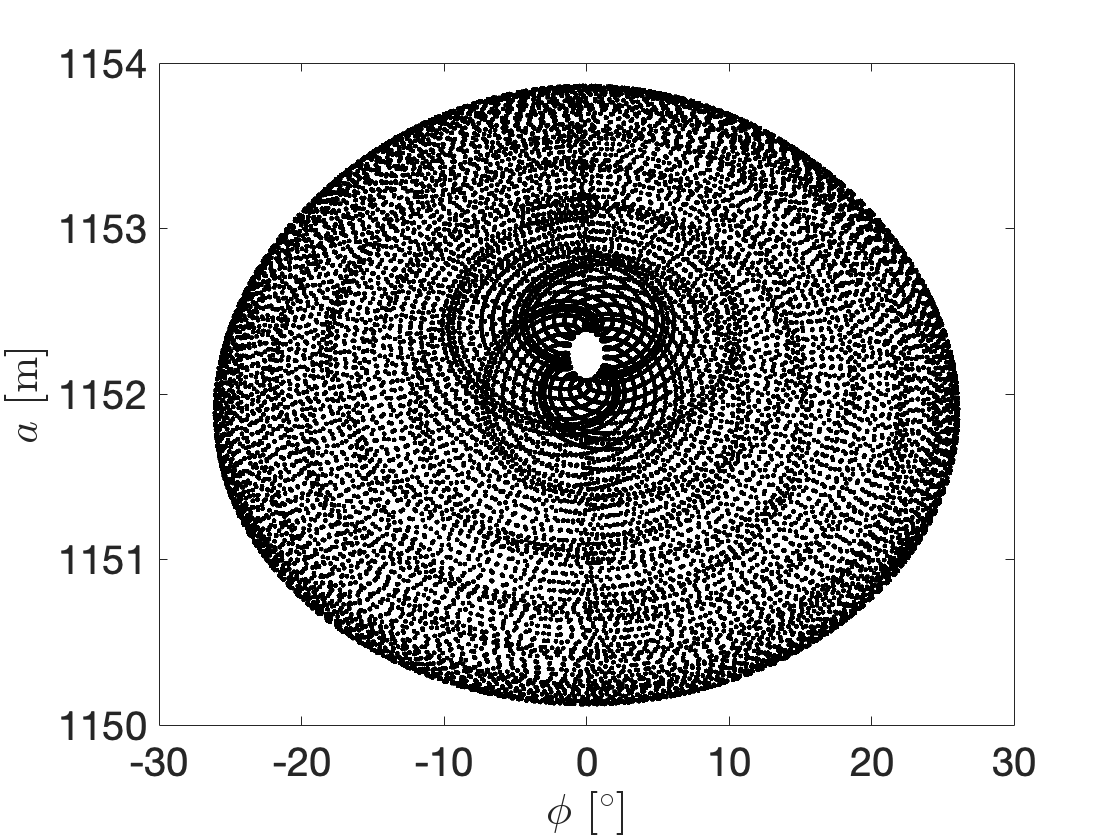}
    \caption{Semimajor axis as a function of libration.}
    \end{subfigure}
    \hfill
    \begin{subfigure}[t]{.5\textwidth}
    \centering
    \includegraphics[width=\linewidth]{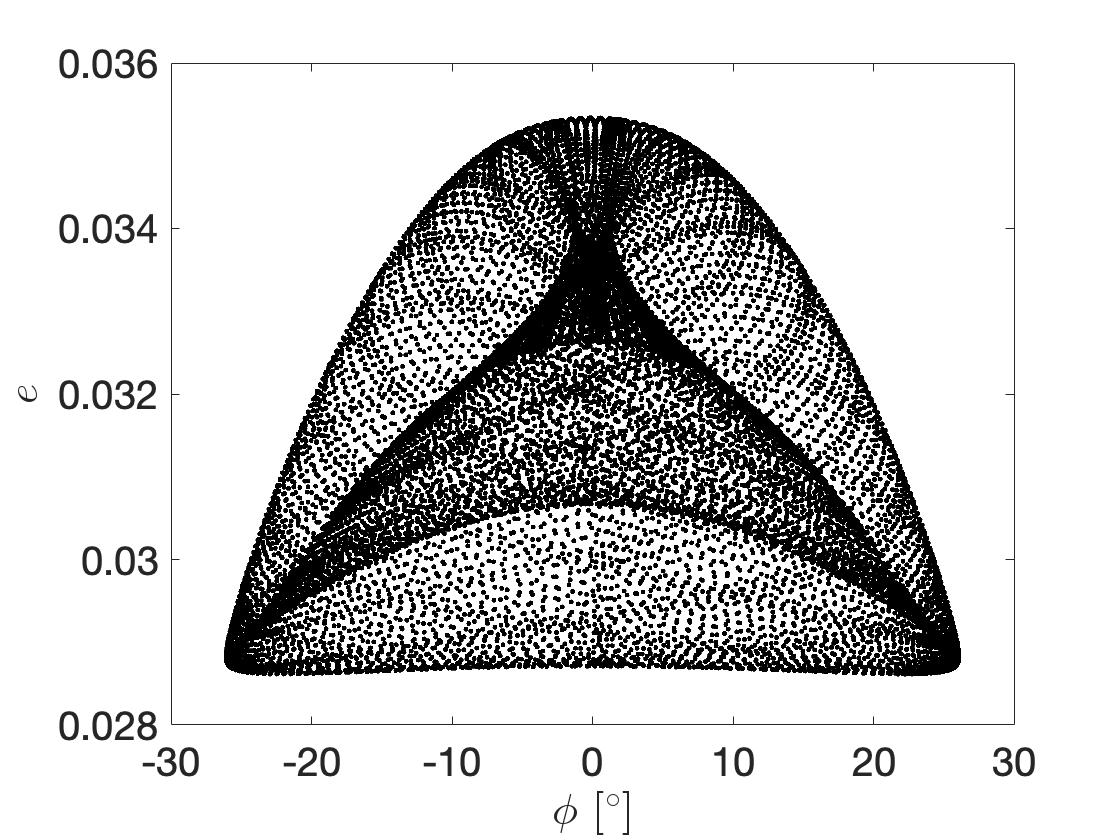}
    \caption{Eccentricity as a function of libration.}
    \end{subfigure}
    
    \caption{Plots from the dynamical model showing the relationship between (a) the libration and true anomaly, (b) separation and libration, (c) semimajor axis and libration, and (d) eccentricity and libration. This illustrates the complicated relationships between these quantities, in particular the importance of the libration angle.}
    \label{fig:parameters}
\end{figure*}

Unfortunately, this expression relies on both $\phi$ and $f$, which are not independent and lack an explicit relationship between the two. Thus, they must be integrated together using proper equations of motion, for example the model in \cite{scheeres2009stability} or \cite{cuk2010orbital}. Furthermore, the semimajor axis, eccentricity, and separation all also depend on these two quantities, and the semimajor axis and eccentricity in particular are not constant. We illustrate the relationships between these parameters in Fig. \ref{fig:parameters}, which shows complex dependencies between true anomaly, libration, separation distance, semimajor axis, and eccentricity.

One may attempt to average Eq. \ref{eq:cuk} over one orbit period. However, as demonstrated in Fig. \ref{fig:parameters}, and also discussed in \cite{meyer2021libration}, there are in reality several frequencies, specifically the frequency of the true anomaly and the frequency of libration. Thus, averaging over one orbit period will not capture the true secular behavior, except for the specific case where the libration frequency is resonant with the orbit mean motion (for a synchronous secondary, this is the case of only forced libration). Alternatively, one may average over one precession period. However, in a purely analytical approach we still run into the problem of the complex relationships between $f$ and $\phi$.

\begin{figure}[ht!]
   \centering
   \includegraphics[width = 3in]{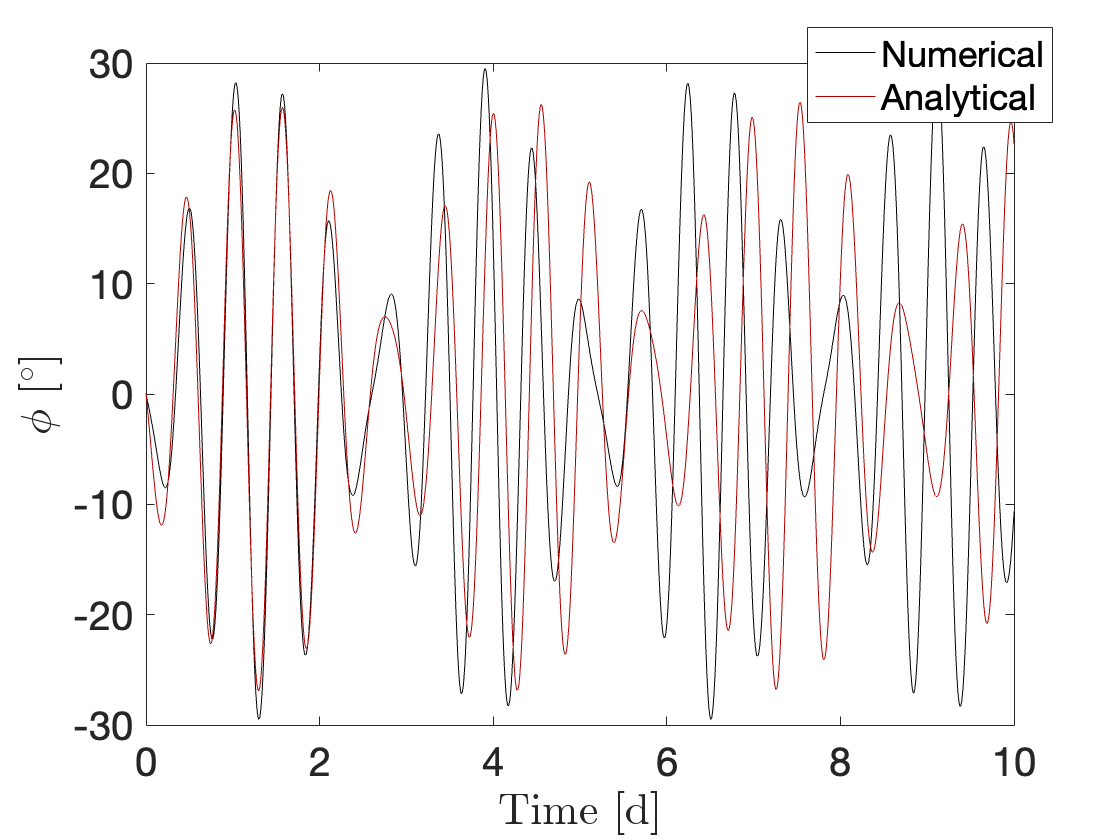} 
   \caption{The time history of the libration angle. The analytical definition diverges from the numerical solution after a few days.}
   \label{fig:libration}
\end{figure}

From Fig. \ref{fig:parameters} it is clear there are several underlying frequencies driving the evolution of the system. These frequencies can be obtained through the linearization of our equations of motion. Specifically, in a system undergoing planar dynamics, it is limited to three degrees of freedom, one of which can be eliminated via the angular momentum integral. This linearization was carried out in \cite{fahnestock2008simulation}, who report the two natural frequencies for a planar system as

\begin{multline}
    \omega_a^2,\omega_b^2 = \bigg[\frac{-12A+3B+9C}{4}-\frac{1}{2}-\frac{3}{2}\nu\kappa\bigg]\frac{9}{4}\pm\\
    \bigg[\frac{9}{4}\bigg(\frac{12A-7B-5C}{2}\nu\kappa\bigg)^2+\\\frac{3}{4}(-20A+17B+3I)-\frac{3}{2}\nu\kappa+\frac{1}{4}\bigg]^\frac{1}{2},
\end{multline}
where $\nu=\frac{m_1}{m_1+m_2}$ and $\kappa=\frac{B-A}{C}$. We can use these frequencies to write an approximation for the true anomaly and libration angle over time. For example, the libration angle can be defined as

\begin{equation}
    \phi = \alpha\cos\omega_at+\beta\cos\omega_bt
\end{equation}
where $\alpha$ and $\beta$ are scalars dependent on the physical parameters of the system. However, as seen in Fig. \ref{fig:libration}, this analytical definition diverges from the numerical solution after only a few days. Thus, for even moderate perturbations from the equilibrium, a linear approximation is inaccurate.

Despite this shortfall, the linearization suggests higher-order approximations will be more accurate in describing the motion within the system. This shows a possible path toward obtaining an analytic averaged expression for the apsidal precession rate in elongated systems. However, even the substitution of the linearized approximation into Eq. \ref{eq:cuk} yields an unwieldy expression containing embedded trigonometric functions, and the derivation is beyond the topic of this work. Thus, we will use a numerical approach, leaving a potential derivation of an accurate, analytically averaged expression to future work.

For a purely analytical approach, we turn to a separate expression for the apsidal precession rate, derived in \cite{jacobson2010orbits}. This model also includes the effects of secondary elongation and libration, but only considers the forced libration within the system. The benefit of reducing the model to only incorporating the forced libration is the relationship between $\phi$ and $f$ becomes much simpler, and we can average over 1 orbit period.  However, unfortunately we cannot use this equation when free libration is present in the system.

For now, we will restrict our dynamics to perfectly planar, so the angular momentum of the secondary and orbit are aligned. Because we do not analytically solve Eq. \ref{eq:cuk}, we will use our dynamical model to calculate the orbital elements and the libration angle. We approximate the apsidal precession rate using a linear fit of the numerically obtained longitude of periapsis. In our dynamics, we start the secondary aligned with the primary ($\phi(0)=0^\circ$) and perturb the velocity to add a small amount of eccentricity to the system. This is similar to the effect of the DART impact in the Didymos binary asteroid system. We test different values of initial $\dot{\phi}$ by adding a spin perturbation $\delta\dot{\phi}$ away from the unperturbed secondary spin rate.

\section{Numerical Approach} \label{sec:numerical}

We use a Didymos-like system to investigate how the apsidal precession rate depends on the system parameters. From the orbital fit derived in \cite{Naidu2024}, we start our system in a mutual circular orbit with a separation distance of $r=1.19$ km. The secondary has a volume-equivalent diameter of 150 m, with semiaxes $a>b>c$ determined by $a/b = 1.3$ and $b/c=1.1$. We assume an equal uniform density of 2.8 g/cm$^3$ between the primary and secondary, and a primary diameter of 710 m. The system equilibrium orbit rate $\dot{\theta}$ is calculated using the expression derived in \cite{scheeres2009stability}:
\begin{equation}
    \dot{\theta}^2 = \frac{\mathcal{G}(m_1+m_2)}{r^3}\left(1+\frac{3}{2r^2}\left(B+C-2A\right)\right).
    \label{eq:spin}
\end{equation}
We set the secondary spin rate equal to this orbit rate for the equilibrium state.

We apply a $\Delta v$ of -3 mm/s to the secondary (tangential and retrograde to the orbital motion), similar to the value caused by the DART impact \citep{cheng2023momentum,Naidu2024}. Along with this velocity perturbation, we also apply a perturbation to the secondary's spin rate $\delta\dot{\phi}$, again consistent with results in \cite{Naidu2024}, who estimate a spin perturbation to Dimorphos. We then integrate the system forward in time for 60 days and calculate the precession rate by linearly fitting to the secular drift in the longitude of periapsis over this time.

\begin{figure*}[ht!]
    \begin{subfigure}[t]{.5\textwidth}
    \centering
    \includegraphics[width=\linewidth]{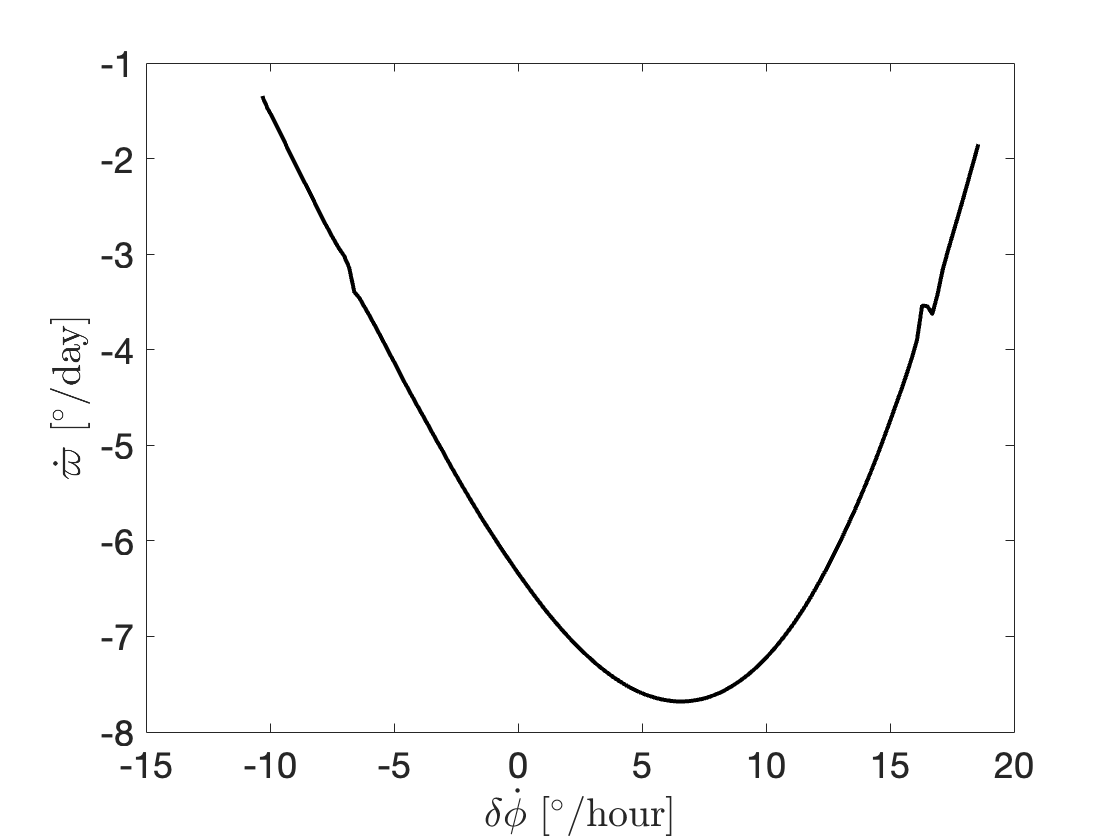}
    \caption{Precession rate as a function of spin perturbation.}
    \end{subfigure}
    \hfill
    \begin{subfigure}[t]{.5\textwidth}
    \centering
    \includegraphics[width=\linewidth]{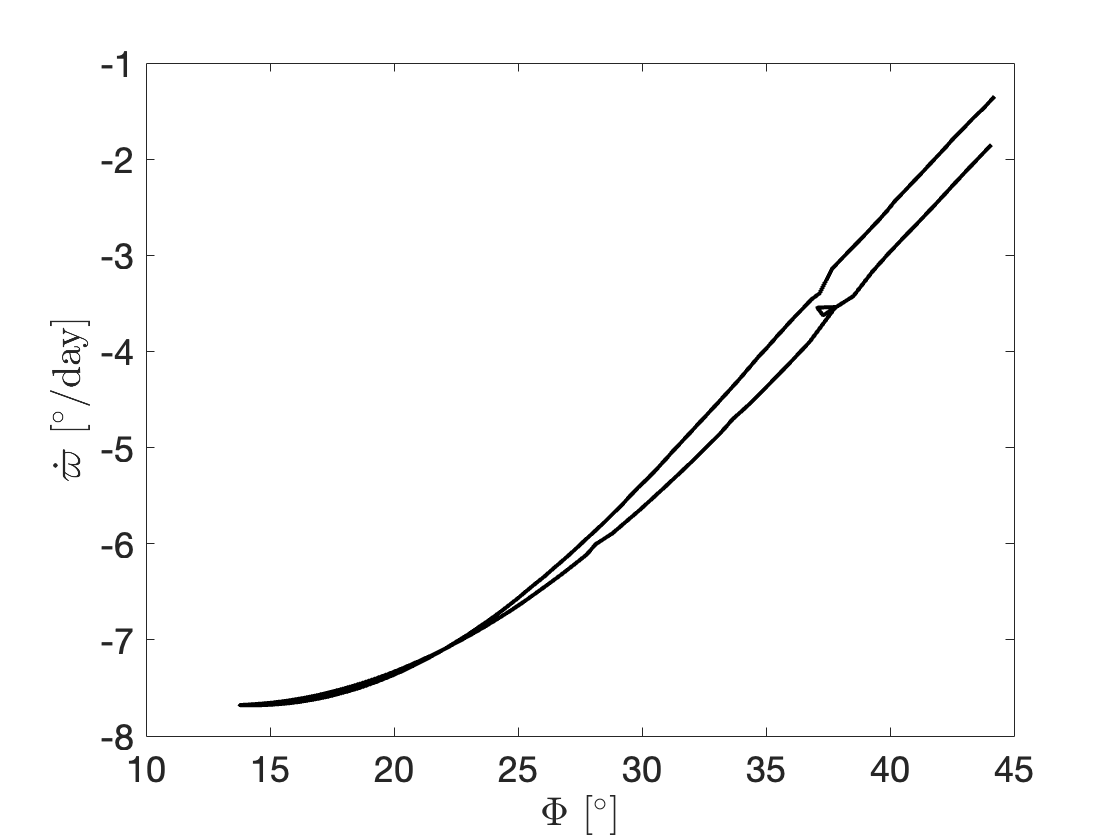}
    \caption{Precession rate as a function of libration amplitude.}
    \end{subfigure}
    
    \medskip
    
    \begin{subfigure}[t]{.5\textwidth}
    \centering
    \includegraphics[width=\linewidth]{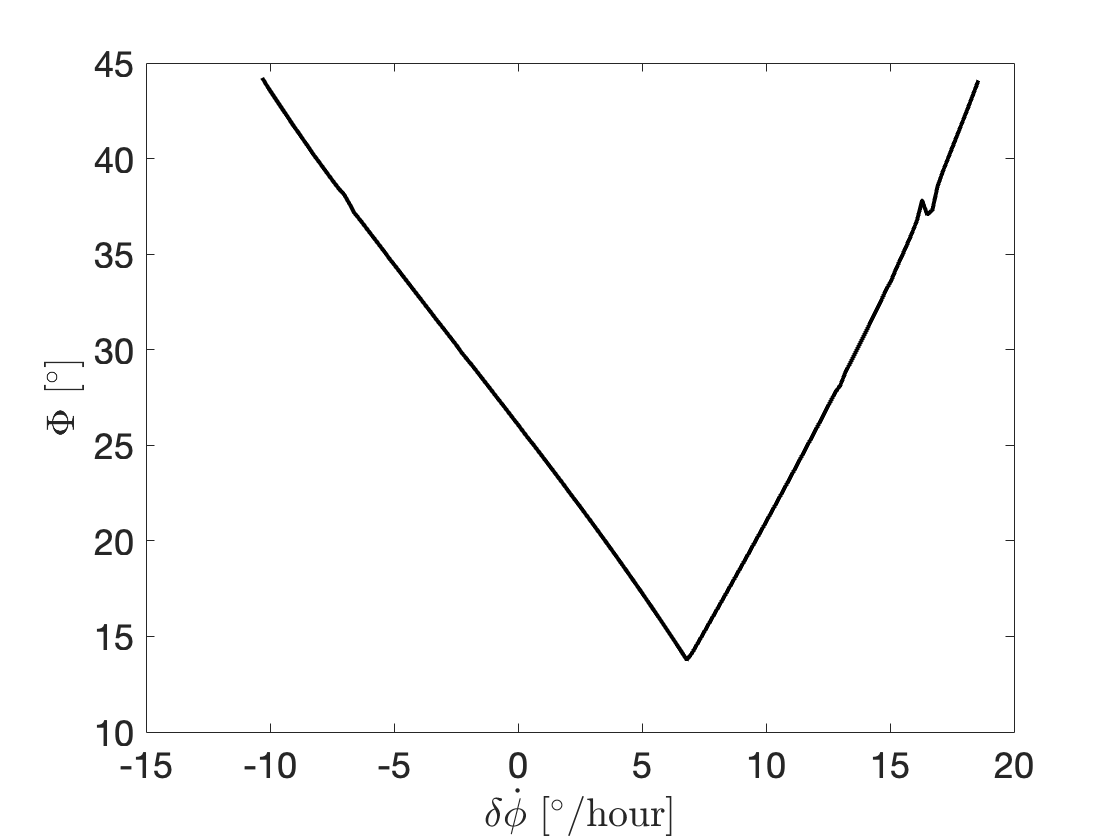}
    \caption{Libration amplitude as a function of spin perturbation.}
    \end{subfigure}
    \hfill
    \begin{subfigure}[t]{.5\textwidth}
    \centering
    \includegraphics[width=\linewidth]{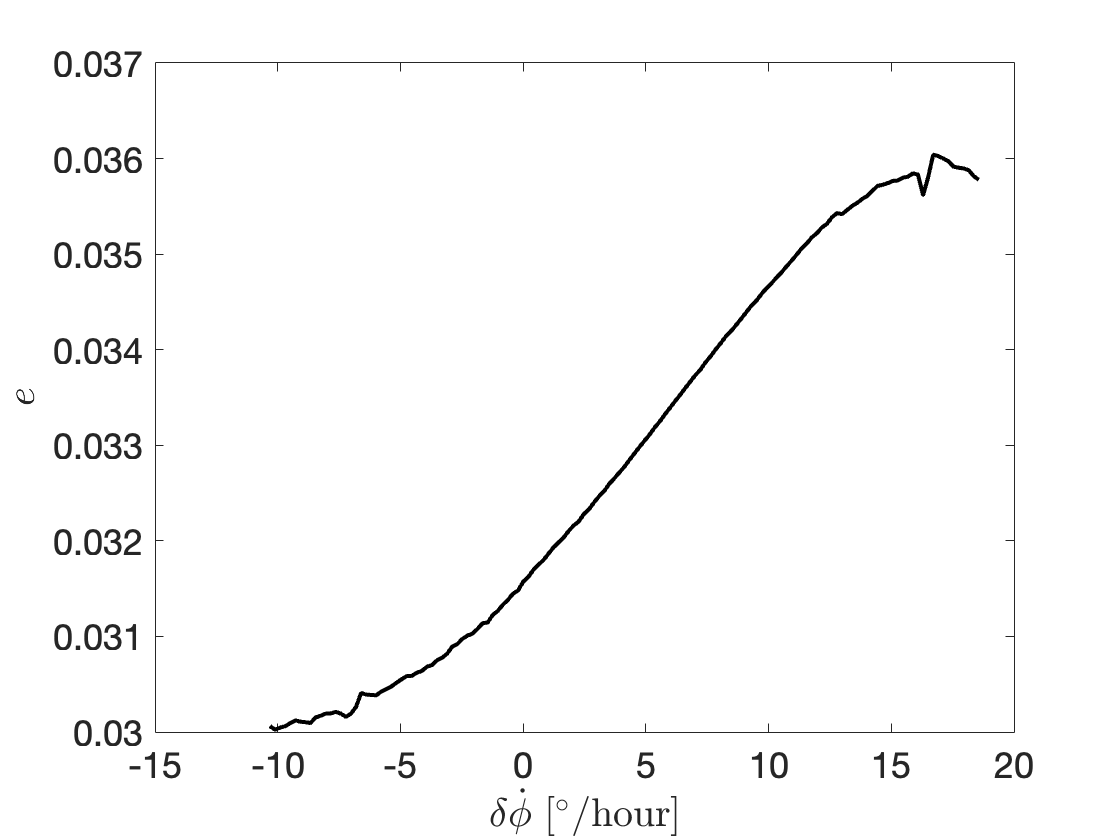}
    \caption{Eccentricity as a function of spin perturbation.}
    \end{subfigure}
    
    \caption{Plots of numeric results showing (a) the precession rate as a function of the spin perturbation, (b) the precession rate as a function of the libration amplitude, (c) the libration amplitude as a function of the spin perturbation, and (d) the eccentricity as a function of the spin perturbation}
    \label{fig:numeric_results}
\end{figure*}

Fig. \ref{fig:numeric_results} shows the results of these simulations. Panel (a) plots the precession rate as a function of the spin perturbation, while panel (b) shows the precession rate as a function of the libration amplitude $\Phi$. The relationship between the libration amplitude and the spin perturbation is shown in panel (c). Finally, panel (d) shows the eccentricity as a function of the spin perturbation. This demonstrates how the precession rate does depend on the secondary's spin, and is not only driven by the changing eccentricity.

From Fig. \ref{fig:numeric_results}, there is a critical spin perturbation which results in the largest apsidal regression. This occurs at the same perturbation which minimizes the libration amplitude, meaning this is the secondary spin rate which eliminates free libration, leaving only the forced libration. Increasing libration amplitude, or equivalently perturbing the secondary spin away from this value, results in decreasing the apsidal regression rate (increasing the precession rate closer to zero). We also see how the secondary's spin perturbation has a small effect on the eccentricity. Increasing the secondary's spin rate results in a higher eccentricity.

In Fig. \ref{fig:numeric_results}, we see some discontinuities in the precession rate for a spin perturbation around 16$^\circ$/hour, and to a lesser extent also around -6$^\circ$/hour. These appear to be due to a resonance between the average orbit period and the average secondary spin period, but they only have a small effect on the dynamics.

\subsection{Comparison with Analytic}

Next we compare these results with the analytic equations for apsidal precession rate. To calculate the precession rate using Eq. \ref{eq:cuk}, we integrate our equations of motion to get a time history for true anomaly and libration angle, then average the time history of $\dot{\varpi}$ to come up with an average apsidal precession rate. In using the forced libration model in \cite{jacobson2010orbits}, we calculate the libration amplitude as the maximum libration angle.

\begin{figure}[ht!]
   \centering
   \includegraphics[width = 3in]{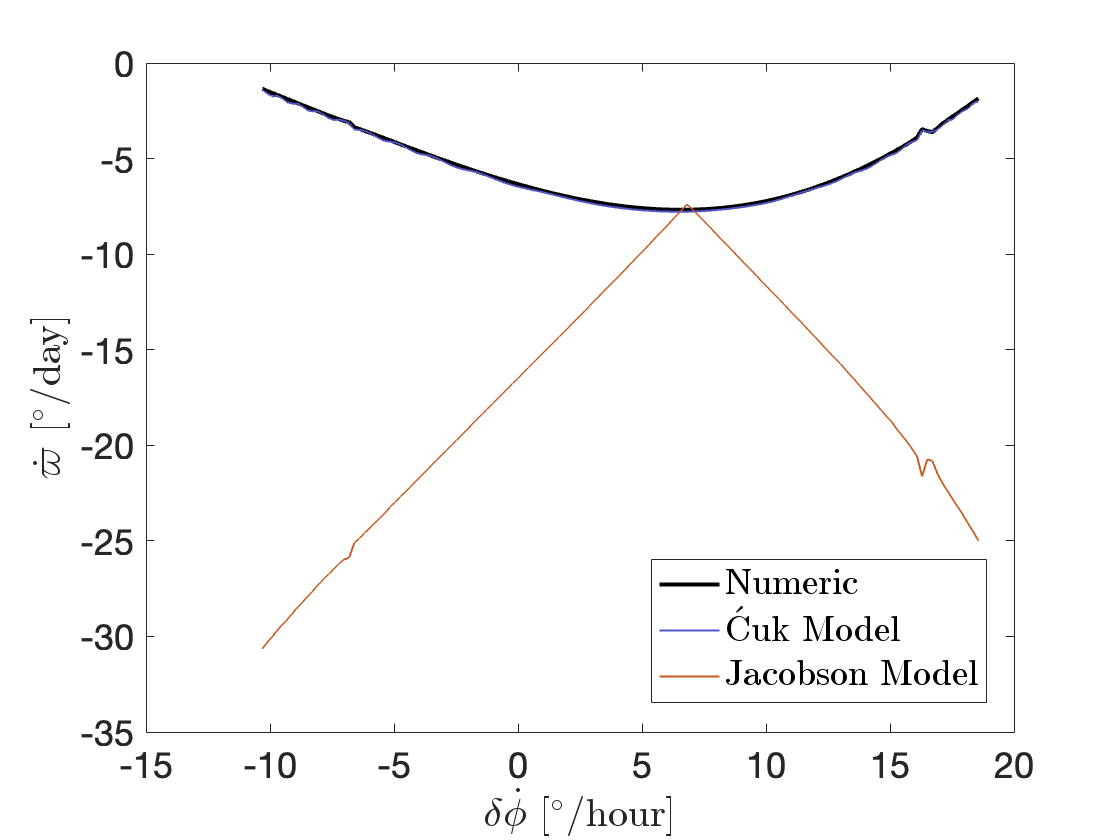} 
   \caption{A comparison between the numerically calculated apsidal precession rate and the analytic expressions. The \'Cuk Model corresponds to Eq. \ref{eq:cuk} and the Jacobson Model corresponds to the result in \cite{jacobson2010orbits}}
   \label{fig:analytic_compare}
\end{figure}

The comparison between these three approaches is shown in Fig. \ref{fig:analytic_compare}. We see  excellent agreement when averaging Eq. \ref{eq:cuk}. The forced libration solution from \cite{jacobson2010orbits} diverges sharply from the numerical solution. This is expected, since this equation only considers the forced libration. When the numerical solution only has forced libration, then the model in \cite{jacobson2010orbits} matches the numerical solution closely. In this system, a spin perturbation around $7^\circ$/hour eliminates free libration from the system, only leaving the forced libration. Around this point, all three models have a close match for the precession rate.

\section{Application to Didymos} \label{sec:didymos}

Next, we apply this analysis to the Didymos system. Using the estimated apsidal precession rate \citep{Naidu2024}, we can place constraints on the secondary shape and libration. We now include the oblateness of the primary, which for Didymos is estimated to be $J_2=0.09$ \citep{Naidu2024}. The estimated apsidal precession rate in Didymos is $\dot{\varpi}=6.7^\circ$/day. The separation of the orbit at the time of the DART impact was 1.189 km, and we assume Dimorphos was in a circular orbit about Didymos prior to the perturbation. We also fix the Dimorphos $b/c=1.2$.

In our algorithm, we iterate a tangential and instantaneous $\Delta v$ on the orbital speed of Dimorphos until our post-impact system matches the measured post-DART orbital period of 11.37 hours. Once this is achieved, we perform a second iteration to the secondary's rotation rate: we add a $\delta \dot{\phi}$ to correctly achieve the estimated apsidal precession rate.

In this work we are exploring the relationship between the secondary shape and spin and the orbit's apsidal precession rate. As a result, we do not account for uncertainties in the size of the two asteroids or in the separation distance at the time of impact. The mass of the asteroids are calculated by inverting Eq. \ref{eq:spin} using the pre-impact orbit period of 11.92 hours, and assuming equal and uniform densities between Didymos and Dimorphos. 

In the system dynamics, the secular apsidal rate is governed by the oblateness of Didymos and Dimorphos, as well as the prolateness of Dimorphos (assuming Didymos to be axially symmetric). The effect of the oblateness is a classical and well understood result. In this work, we focus only on the role of the prolateness. 

\begin{figure}
    \begin{subfigure}[t]{.5\textwidth}
    \centering
    \includegraphics[width=\linewidth]{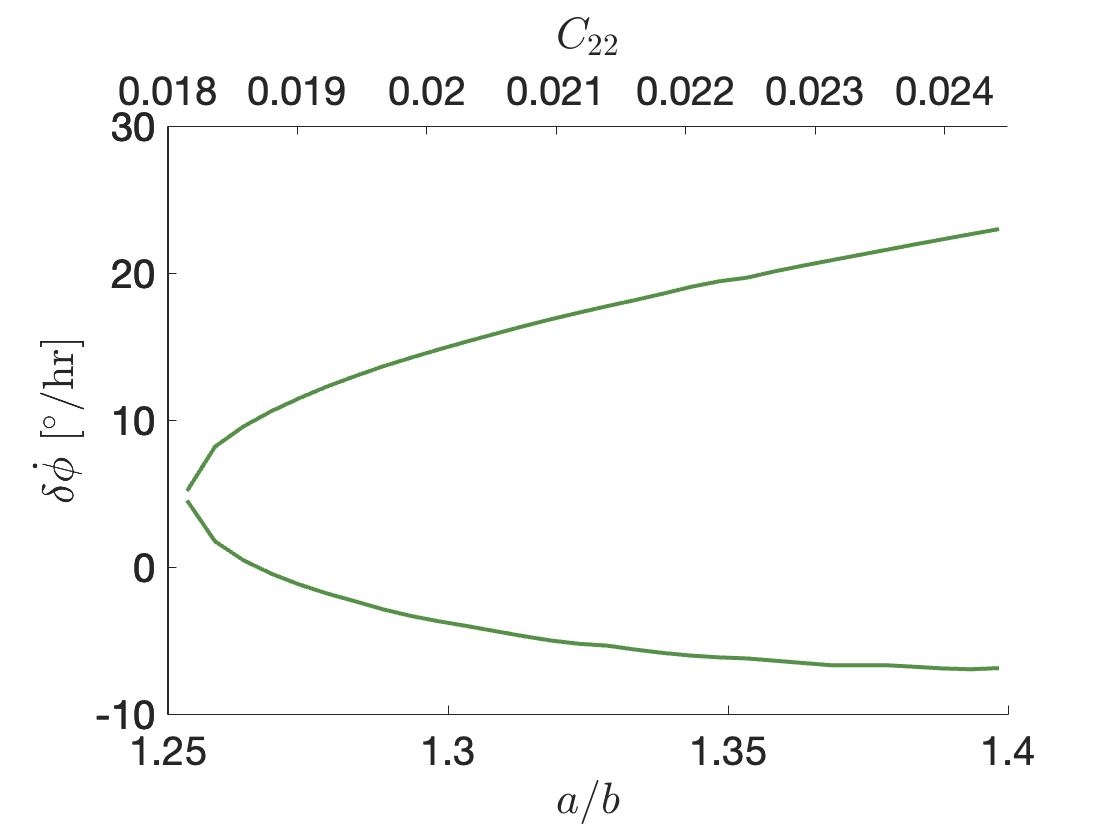}
    \caption{Spin perturbation as a function of prolateness.}
    \label{fig:didy_pert_v_ab}
    \end{subfigure}

    \begin{subfigure}[t]{.5\textwidth}
    \centering
    \includegraphics[width=\linewidth]{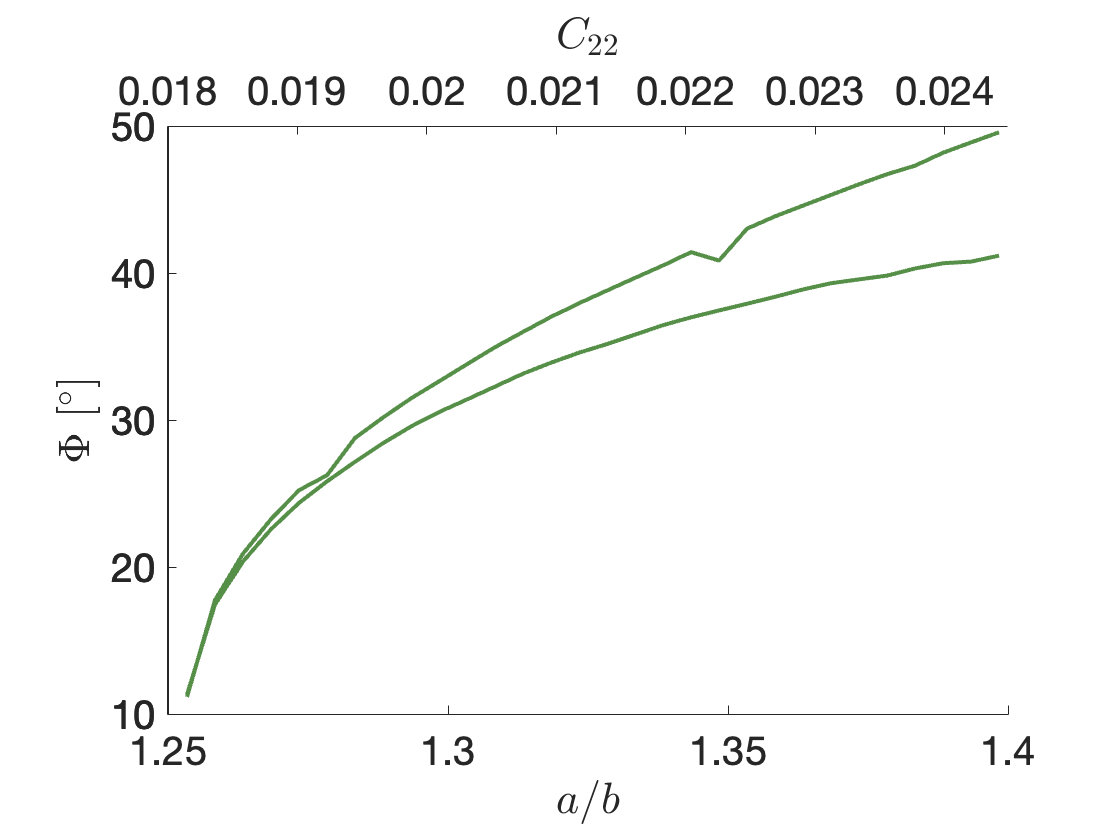}
    \caption{Libration amplitude as a function of prolateness.}
    \label{fig:didy_lib_v_ab}
    \end{subfigure}

    \begin{subfigure}[t]{.5\textwidth}
    \centering
    \includegraphics[width = \linewidth]{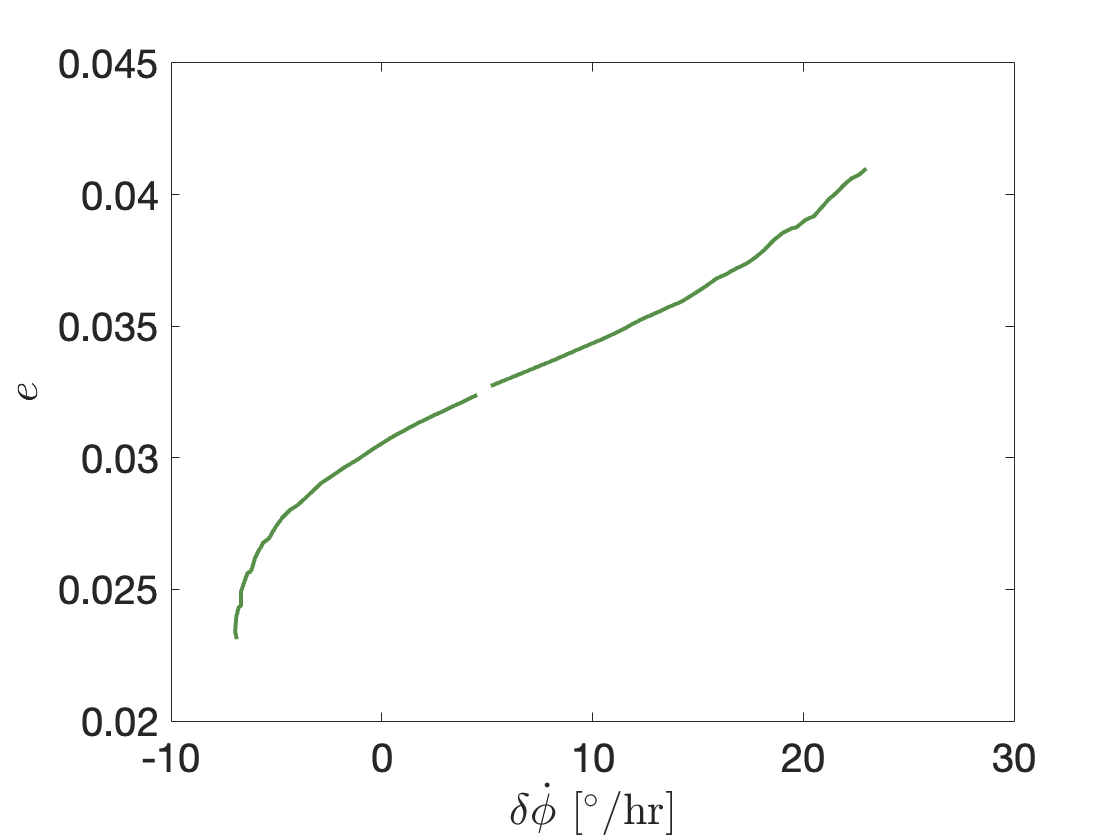} 
    \caption{Eccentricity as a function of spin perturbation}
    \label{fig:didy_e_v_pert}
    \end{subfigure}
    
    \caption{Plots of numeric results showing (a) The required spin perturbation, (b) the required libration amplitude as functions of Dimorphos prolateness necessary to reach the measured precession rate, and (c) the resulting eccentricity as a function of the spin perturbation at the measured precession rate.}
\end{figure}

In Fig. \ref{fig:didy_pert_v_ab}, we plot curves that represent combinations of secondary elongation and spin perturbation that replicate the apsidal precession rate in Didymos. In Fig. \ref{fig:didy_lib_v_ab}, we plot the same curves but for libration amplitude. For a given Dimorphos $C_{22}$, or equivalently an $a/b$ value, there are two values of spin perturbation $\delta\dot{\phi}$ that can reproduce the observed precession rate. Thus, some knowledge of the perturbation's geometry is required to resolve the ambiguity between the two curves. For the geometry of the DART impact, we likely have a negative value of $\delta\dot{\phi}$ \citep{Naidu2024}, corresponding to the lower of the two curves in Figs. \ref{fig:didy_pert_v_ab} and \ref{fig:didy_lib_v_ab}.

These curves demonstrate that for the nominal system (ignoring uncertainties on the shape of Didymos and the orbit geometry), we are required to have an $a/b\gtrsim1.25$, equivalent to having $C_{22}\gtrsim0.018$. This is a purely dynamical constraint indicating likely significant reshaping in Dimorphos as a result of the DART impact, which complements the physical argument for the same phenomenon by \cite{raducan2024physical}. This is also consistent with the estimation of the system parameters by \cite{Naidu2024}.

However, we have only used the nominal system parameters. In reality, smaller values of $a/b$ are permitted, for example if the primary's $J_2$ term is in reality smaller than the best estimate. Taking into account all uncertainties, it is possible for the pre-impact and post-impact shape of Dimorphos to be essentially the same and still produce the measured precession rate. However, this case requires extreme values for the Didymos $J_2$ and the separation distance, which is a fringe case. Thus, we argue it is very likely Dimorphos experienced some degree of reshaping, with its new shape more elongated than the pre-impact shape.

We can also use this approach to investigate the allowable libration amplitude in the system. Here we see uniformly increasing libration amplitude with increasing $C_{22}$. The minimum $C_{22}$ results in a libration amplitude of around $10^\circ$, indicating the nominal system has at least this libration amplitude. Again, smaller values are permissible given the uncertainties in the system, but most combinations of parameters will result in a libration amplitude larger than this value.

As we have already seen, the eccentricity is also affected by the spin perturbation. In Fig. \ref{fig:didy_e_v_pert}, we see increasing eccentricity with increasing $\delta\dot{\phi}$. This illustrates the spin-orbit coupling inherent to binary asteroids: as the angular momentum of the secondary changes, this is reflected in the angular momentum of the orbit as well.

\section{Out of Plane Rotation} \label{sec:npa}

Next we examine the role played by out of plane rotation on the orbital precession. Recall our dynamical model permits out of plane rotation by the secondary. It has been well established that resonances within the binary asteroid system can induce attitude instabilities in the secondary \citep{agrusa2021excited}. Typically, the attitude of the secondary is reported as a set of 1-2-3 Euler angles corresponding to its roll, pitch, and yaw, relative to a rotating hill frame \citep{meyer2021effect,agrusa2021excited,richardson2022predictions}, and we will adopt this convention as well. 

We are interested in how the apsidal precession rate is affected by non principal-axis (NPA) rotation, particularly applied to Didymos. Thus, we keep our same algorithm as described in Section \ref{sec:didymos}, but now also perturb the secondary's out of plane rotation. When we calculate the spin perturbation about the major principal axis to achieve the estimated precession rate in the planar case, we apply an additional perturbation no larger than this value to the other axes as well.

In their work, \cite{Pravec2024} break up NPA rotation of the binary secondary into three regimes: No tumbling, epicyclic tumbling, and constant tumbling. We show an example from each of these regimes, demonstrating the effect of NPA rotation on the apsidal precession rate.

\begin{figure}[ht!]
    \begin{subfigure}[t]{.5\textwidth}
    \centering
    \includegraphics[width=\linewidth]{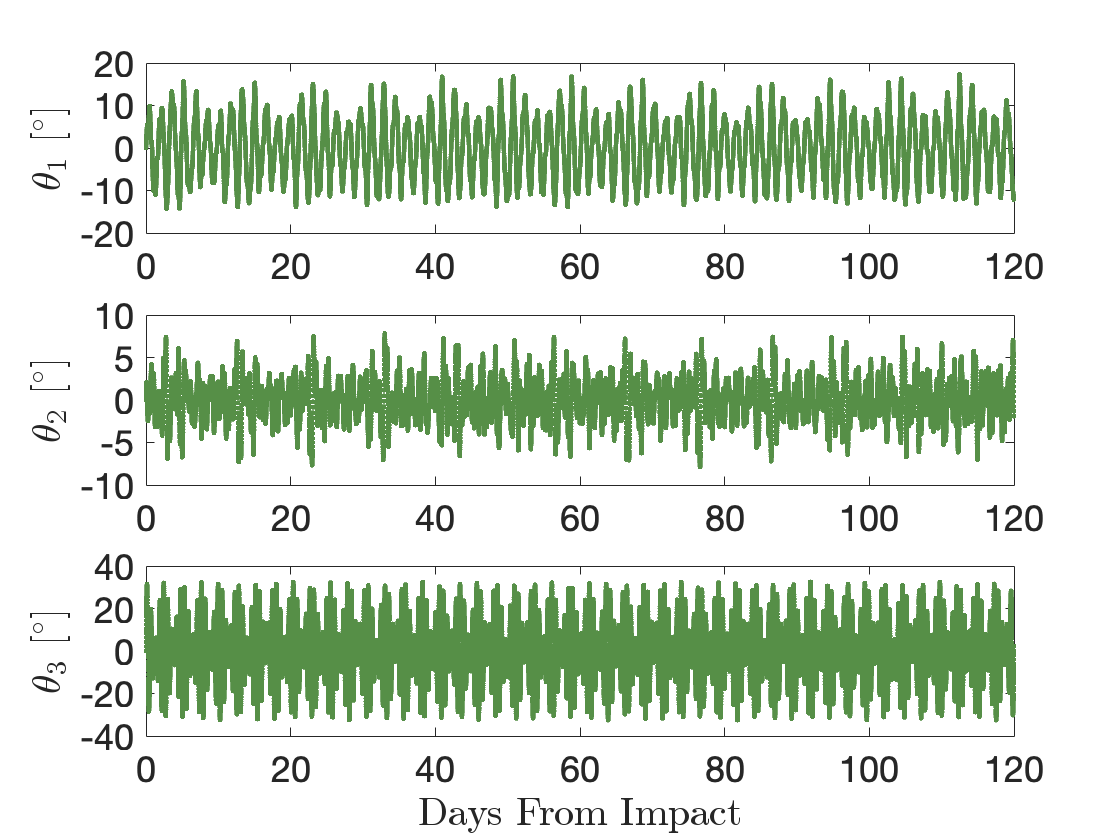}
    \caption{1-2-3 Euler Angles.}
    \end{subfigure}

    \begin{subfigure}[t]{.5\textwidth}
    \centering
    \includegraphics[width=\linewidth]{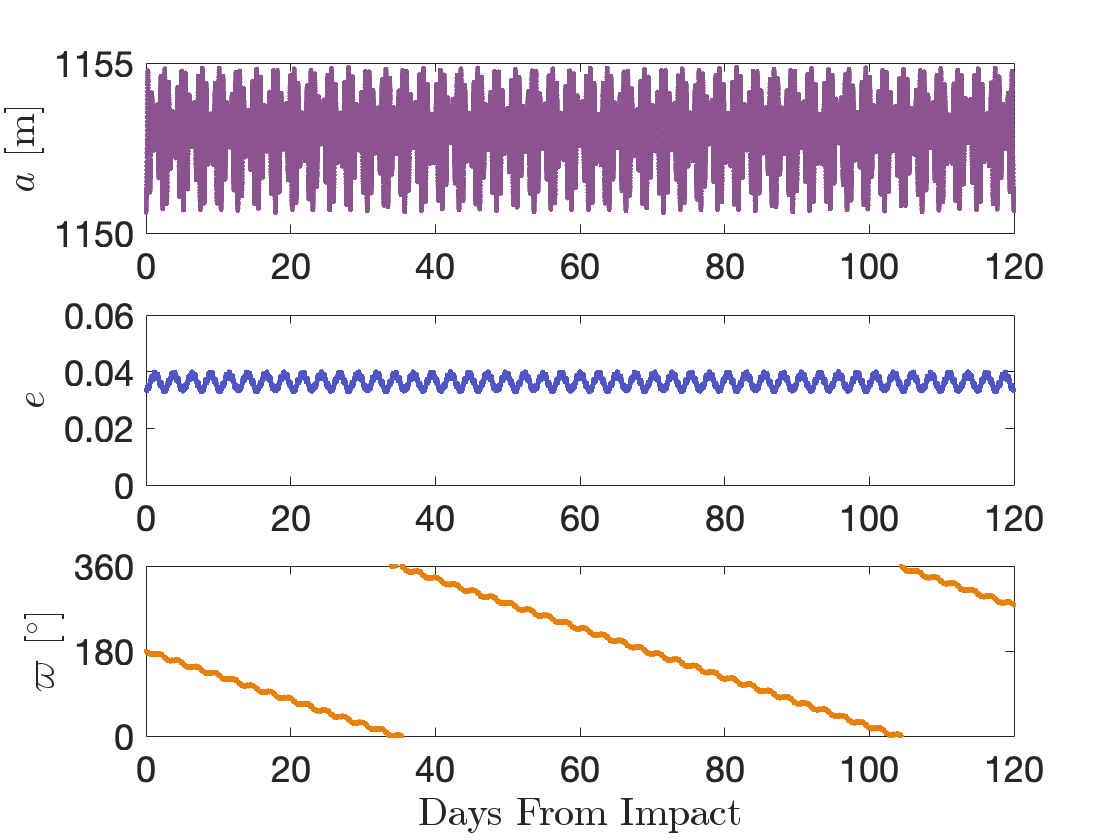}
    \caption{True anomaly, longitude of periapsis, and eccentricity.}
    \end{subfigure}

    \begin{subfigure}[t]{.5\textwidth}
    \centering
    \includegraphics[width=\linewidth]{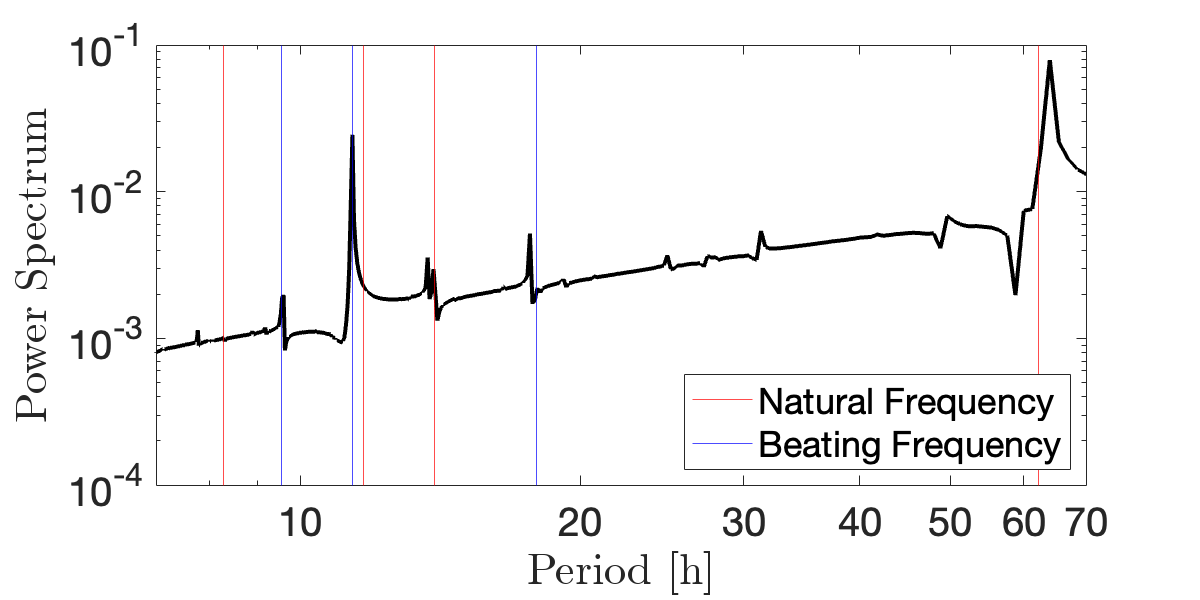}
    \caption{Power spectrum of $e\cos\varpi$.}
    \end{subfigure}
    
    \caption{Illustration of no tumbling in NPA rotation (stable), characterized by small out of plane Euler angles, a constant precession rate, and relatively constant eccentricity. The power spectrum shows the influence of the natural frequencies, as well as several beating frequencies. This demonstrates the motion is quasiperiodic}.
    \label{fig:regime1}
\end{figure}

When NPA rotation is very minimal, we essentially have no tumbling in the secondary. In this regime, the out of plane Euler angles are small and quasiperiodic, and the system behaves nearly identically to principal axis rotation. An example of NPA rotation with no tumbling is shown in Fig. \ref{fig:regime1}. We see generally well-behaved dynamics. The NPA rotation is small, and as a result we see nearly constant semimajor axis and eccentricity, and an on-average constant apsidal precession.

Fig. \ref{fig:regime1} also shows the power spectrum of $e\cos\varpi$. We see the influence of the natural frequencies (now also including the two additional natural frequencies from \cite{fahnestock2008simulation}). Beating between these natural frequencies is also important as illustrated in Fig. \ref{fig:regime1}. The dominance of these natural frequencies within the power spectrum indicates the motion is still quasiperiodic despite the presence of NPA rotation in the no tumbling regime. If high fidelity measurements are able to detect these frequencies within the data, they can provide information about the physical parameters of the system.

\begin{figure}[ht!]
    \begin{subfigure}[t]{.5\textwidth}
    \centering
    \includegraphics[width=\linewidth]{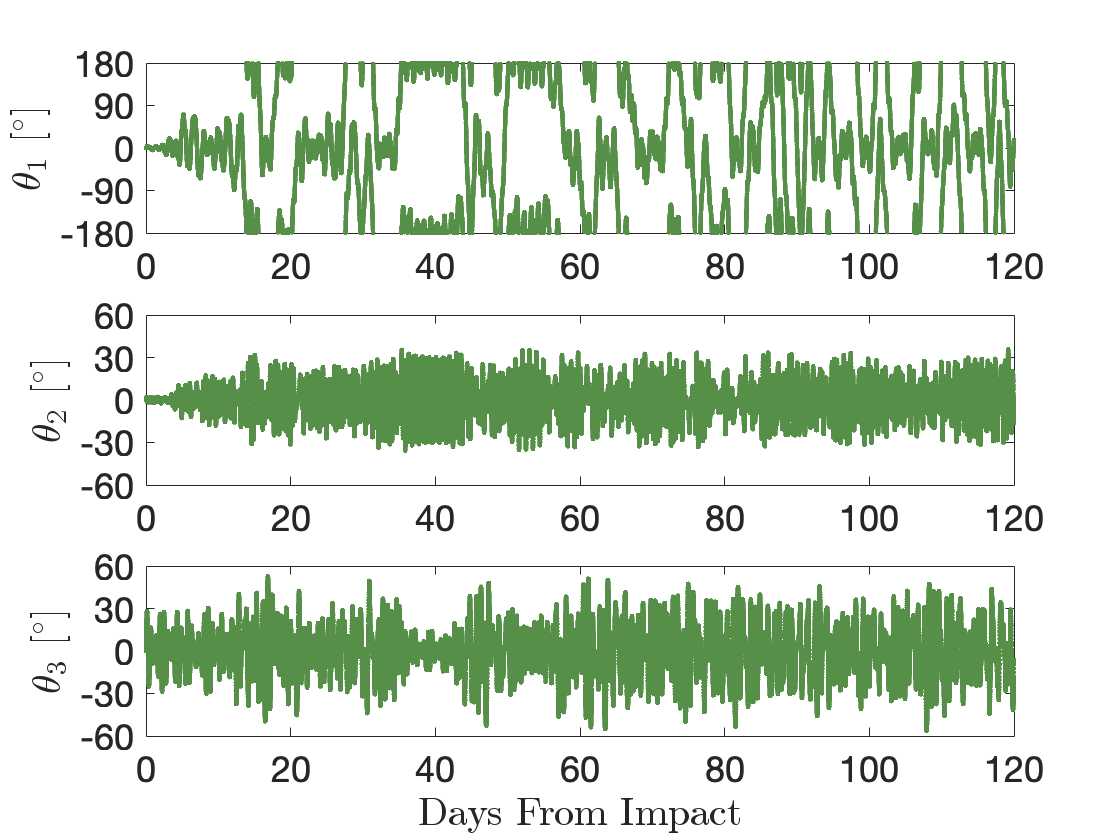}
    \caption{1-2-3 Euler Angles.}
    \end{subfigure}

    \begin{subfigure}[t]{.5\textwidth}
    \centering
    \includegraphics[width=\linewidth]{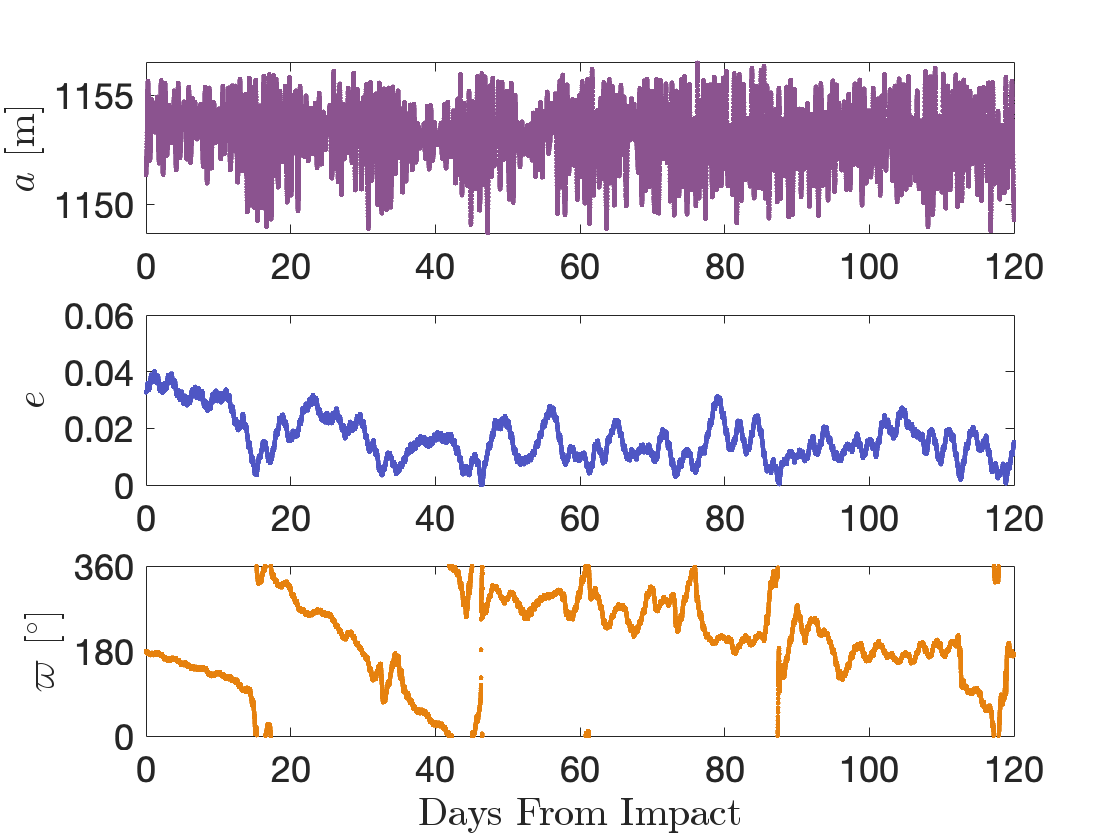}
    \caption{True anomaly, longitude of periapsis, and eccentricity.}
    \end{subfigure}

    \begin{subfigure}[t]{.5\textwidth}
    \centering
    \includegraphics[width=\linewidth]{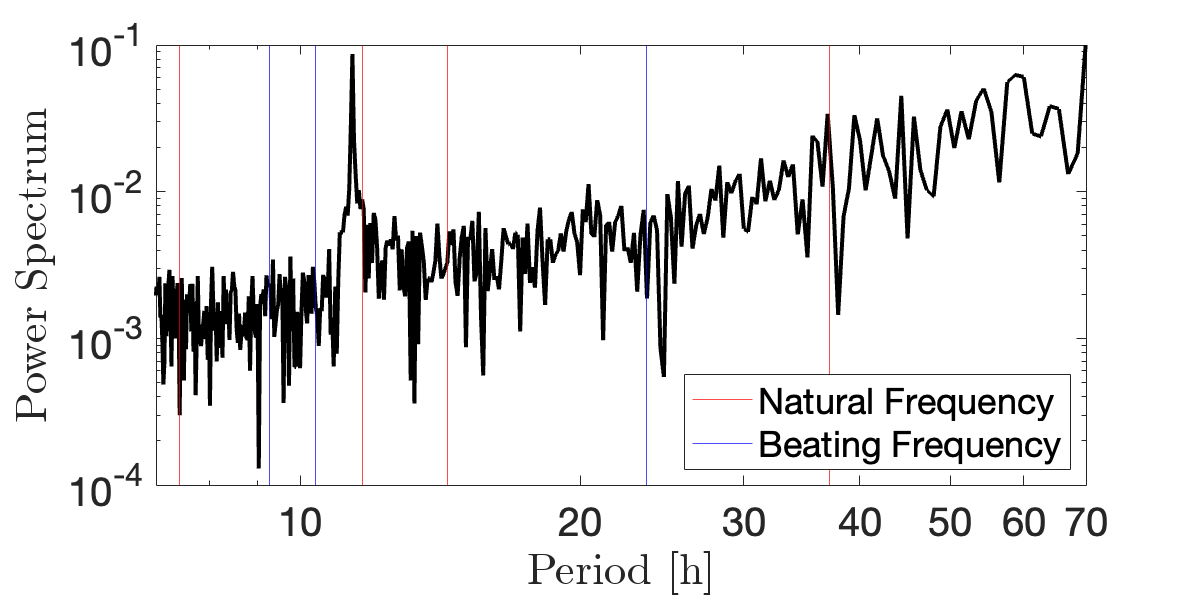}
    \caption{Power spectrum of $e\cos\varpi$.}
    \end{subfigure}
    
    \caption{Illustration of epicyclic tumbling in NPA rotation (synchronous tumbling), characterized by large out of plane Euler angles with $\theta_3$ rotating around 0 or 180$^\circ$, occasional rapid increases in the precession rate, and rapid changes in the eccentricity. The power spectrum shows the motion is not longer quasiperiodic, as the natural frequencies no longer dominate}.
    \label{fig:regime2}
\end{figure}

In epicyclic tumbling, the dynamics become more complicated as out of plane rotation angles can become large, and the secondary frequently enters the barrel instability \citep{cuk2021barrel}. However, generally the secondary remains on-average synchronous, where its long axis points toward or away from the primary. In this regime, the libration amplitude varies as angular momentum is exchanged between the secondary and the orbit, and as a result the apsidal precession rate can vary as well. An example of epicyclic tumbling is shown in Fig. \ref{fig:regime2}. The barrel instability begins around 20 days into the simulation and is shown by `flipping' of the roll angle. As angular momentum is exchanged between the secondary and orbit, we see a corresponding drop in eccentricity, as explained in \cite{meyer2023perturbed}. The semimajor axis sees a small amount of variation, and the apsidal precession rate can fluctuate significantly away from the average rate, even reversing occasionally. Since the eccentricity is not constant, we see corresponding changes to the apsidal precession rate as expected.

The power spectrum for epicyclic tumbling is also shown in Fig. \ref{fig:regime2}. Now we see the loss of prominance for the natural frequencies within the system, with the exception of the mean motion. Thus, the motion in epicyclic tumbling is no longer quasiperiodic and becomes chaotic. Thus, connecting the frequencies to physical parameters in the system becomes impossible once tumbling has begun.

\begin{figure}[ht!]
    \begin{subfigure}[t]{.5\textwidth}
    \centering
    \includegraphics[width=\linewidth]{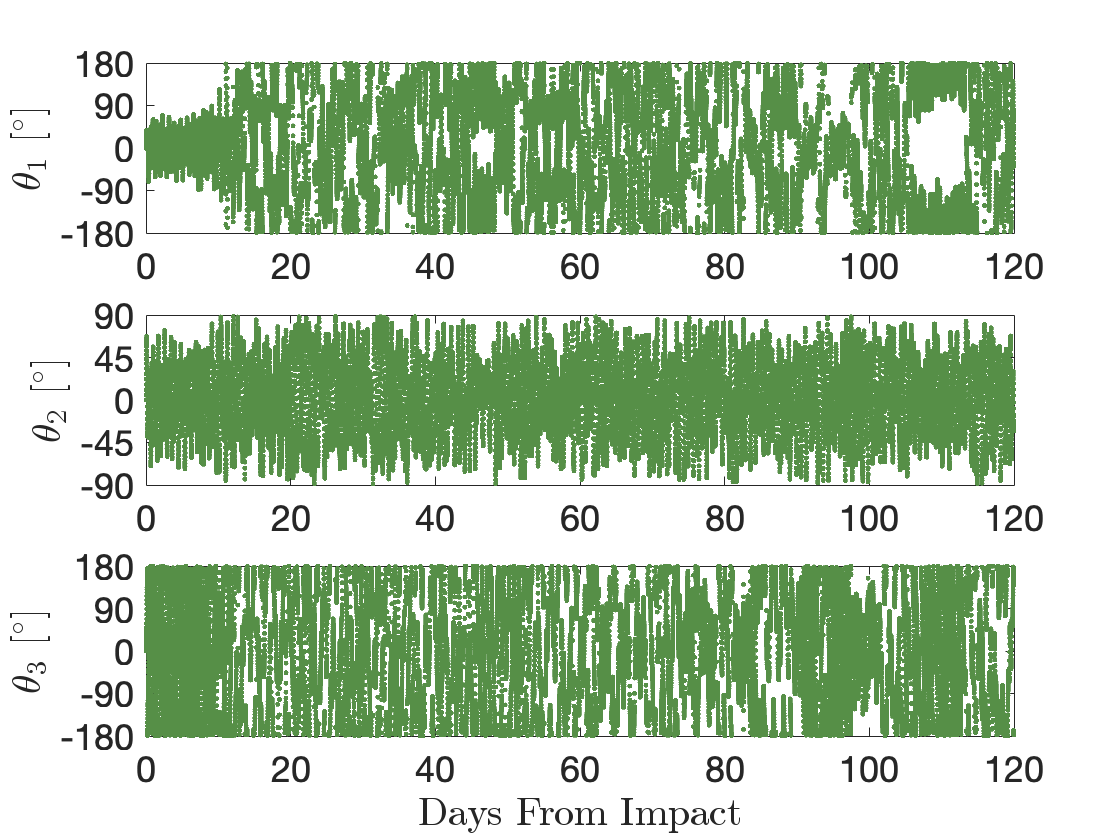}
    \caption{1-2-3 Euler Angles.}
    \end{subfigure}

    \begin{subfigure}[t]{.5\textwidth}
    \centering
    \includegraphics[width=\linewidth]{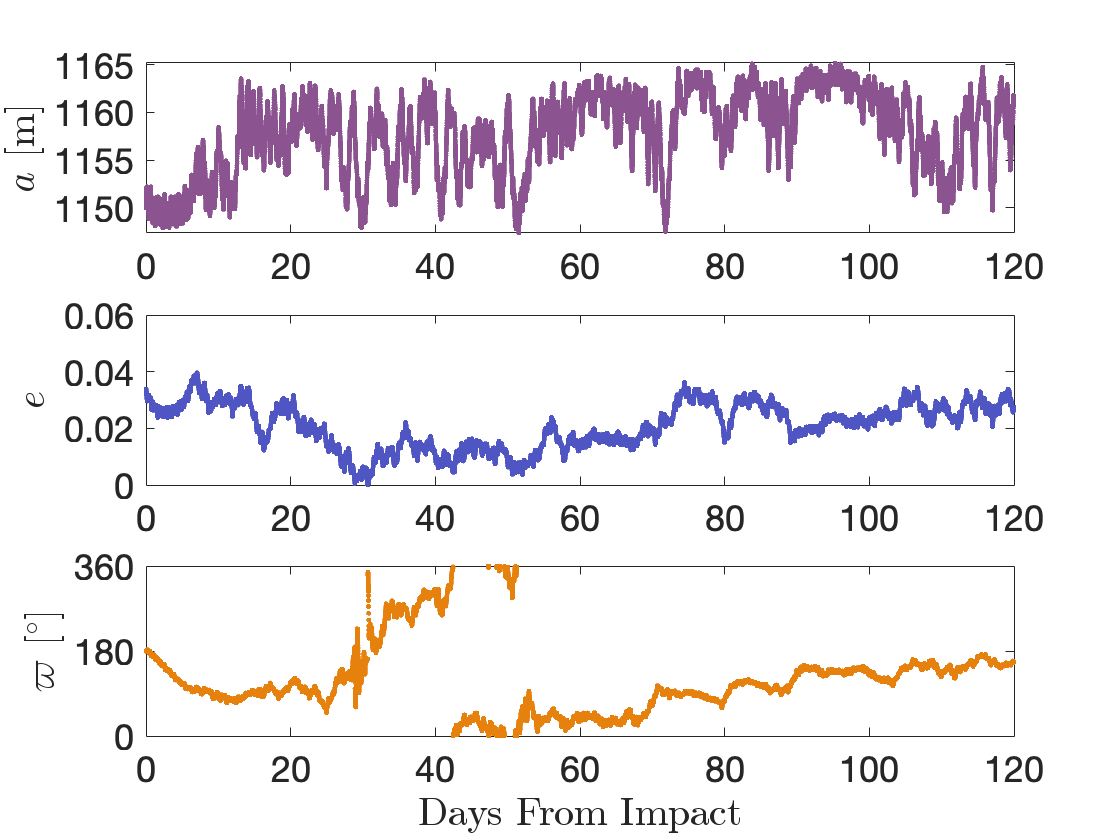}
    \caption{True anomaly, longitude of periapsis, and eccentricity.}
    \end{subfigure}

    \begin{subfigure}[t]{.5\textwidth}
    \centering
    \includegraphics[width=\linewidth]{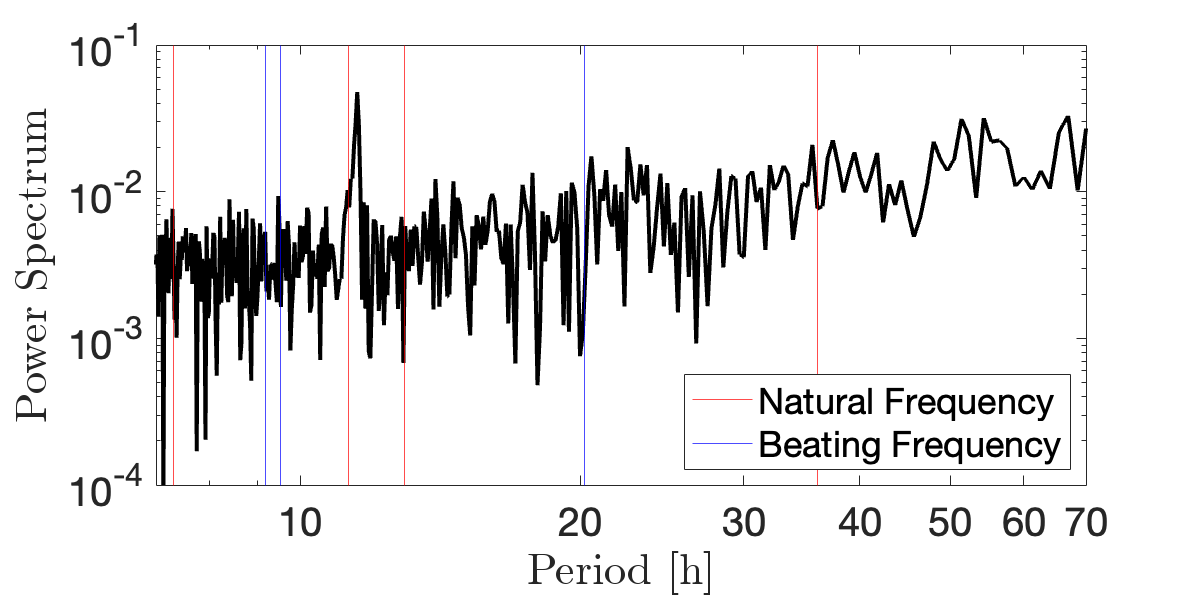}
    \caption{Power spectrum of $e\cos\varpi$.}
    \end{subfigure}
    
    \caption{Illustration of constant tumbling in NPA rotation (chaotic tumbling), characterized by large and uncorrelated out of plane Euler angles, and relatively constant precession rate and eccentricity. The power spectrum shows the motion is not longer quasiperiodic, as the natural frequencies no longer dominate}.
    \label{fig:regime3}
\end{figure}

Finally, \cite{Pravec2024} define constant tumbling as fully asynchronous chaotic rotation of the secondary. An example of this is shown in Fig. \ref{fig:regime3}. Here we see no real structure to the Euler angles, and all 3 see significant amplitudes. This corresponds to fully chaotic secondary rotation. The semimajor axis sees more variation, along with the eccentricity. Interestingly, the longitude of periapsis does not have much of a secular drift, but varies with the eccentricity. Again, the power spectrum in Fig. \ref{fig:regime3} reveals chaotic motion within the system, as the natural frequencies do not have a strong influence on the power spectrum outside the mean motion.

From these examples it is evident that NPA rotation of the secondary has a substantial impact on the apsidal precession of the orbit. While rotation within the no tumbling regime does not have much of an effect, epicyclic tumbling can cause significant departures from the expected precession rate, and constant tumbling can destroy this secular drift altogether. Thus, the extant estimate of Didymos's precession rate over the first $\sim70$ days after the DART impact suggests Dimorphos was in planar rotation during this time period. However, after this time, NPA rotation is likely as detailed in \cite{Pravec2024}.

\section{Discussion} \label{sec:discussion}
In this work we have investigated the role played by the secondary in the secular evolution of the orbit in a binary asteroid. The apsidal precession rate of the orbit is driven by the elongation and spin rate of the secondary, as well as the orbit geometry and mean motion. Thus, unravelling the relationship between observations of the apsidal precession rate and the physical system parameters is a complicated problem and we are limited to numerical investigation.

We find a critical spin rate of the secondary results in eliminating free libration in the system. This is an expected result, but has not been investigated in binary asteroids previously. This critical spin rate places restrictions on the physical parameters of the system given an estimated precession rate. For the nominal estimates of primary $J_2$, separation distance, and apsidal precession rate in Didymos after the DART impact, we can constrain Dimorphos's elongation to $a/b\gtrsim1.25$, with a corresponding minimum libration amplitude of about $10^\circ$. Furthermore, a negative spin perturbation serves to decrease the eccentricity of the orbit after the DART impact. This can explain the difference between the eccentricity estimates obtained by \cite{meyer2023perturbed} who did not account for this parameter and the smaller estimate given in \cite{Naidu2024}. 

Our constraint on Dimorphos's nominal elongation is more restrictive than the values obtained in the analysis by \cite{Pravec2024}, who find possible secondary elongations as low as $a/b=1.1$. In this work we only focused on the nominal system parameters, and did not account for uncertainties in the orbit semimajor axis, Didymos $J_2$, the apsidal precession rate, or Dimorphos $b/c$. Thus, smaller elongations and libration amplitudes are in reality possible when accounting for uncertainty in $J_2$ and the semimajor axis. However, we are mainly concerned with unravelling the relationship between different observable parameters within the binary asteroid. But from the nominal system parameters, we expect significant reshaping in Dimorphos, consistent with results obtained from impact simulations \citep{raducan2024physical}. These results are also all consistent with the best fit of the post-impact dynamical state of Didymos obtained by \cite{Naidu2024}

Given the dynamically coupled nature of binary asteroids, NPA rotation in the secondary can also affect the orbital precession rate. While a small amount of NPA rotation does not have an appreciable effect, larger amplitudes of  out of plane rotation can remove the secular trend in the longitude of periapsis altogether. Power spectrum decompositions of our simulations demonstrate that if a system is not tumbling, the natural frequencies within the system dominate and we can use these to constrain the physical parameters within the system. For Didymos, this is how we are able to argue reshaping has occurred in Dimorphos. However, once NPA rotation becomes significant the natural frequencies within the data no longer drive the evolution of the system. The benefit of this is that the loss of these signals indicates the presence of tumbling in the secondary. Thus, the loss of the precession signal around 70 days after the DART impact as reported in \cite{scheirich2024dimorphos} is an indication of tumbling in Dimorphos. Unfortunately, this means leveraging the natural frequencies or precession period becomes impossible once tumbling has begun.

So far we have only used the apsidal precession signal, but the direct detection of the natural frequencies could provide further constraints to the system. Unfortunately, these short-period frequencies are difficult to detect, and could possibly be dominated by the mean motion. It would require very high quality data to detect and separate these frequencies, likely impossible using ground-based observations.

We have demonstrated that the apsidal precession rate is an important observable within binary asteroids that can help to constrain the physical parameters of the system. However, the relationship between these is complicated and can be ambiguous. There are a range of combinations between $J_2$, $a/b$, $b/c$, and $\delta\dot{\phi}$ that can produce a measured precession rate. Furthermore, the problem is made more complicated by the presence of NPA rotation, which can have a significant effect on the orbital precession rate.

\section{Conclusion} \label{sec:conclusion}

We have investigated how the physical and dynamical properties of a binary asteroid secondary affect the apsidal precession rate of its mutual orbit. Owing to complex relationships in these spin-orbit coupled systems, we relied primarily on numerical simulations. These demonstrated how spin perturbations to the secondary have a significant affect on the precession rate. When applied to Didymos and the DART impact, we found a lower limit on the possible elongation of Dimorphos that is consistent with the estimated precession rate, suggesting significant reshaping of the asteroid. This places constraints on the post-impact shape and spin state of the secondary, pointing a libration amplitude likely larger than $10^\circ$. Finally, we found NPA rotation in the secondary has a major impact on the orbit's precession rate.

\bibliography{bib}{}
\bibliographystyle{aa}

\end{document}